\documentclass{svproc}
\usepackage{cite}
\usepackage{amsmath,amssymb,amsfonts}
\usepackage{algorithm}
\usepackage{algorithmicx}
\usepackage{algpseudocode}
\usepackage{graphicx}
\usepackage{textcomp}
\usepackage{xcolor}
\usepackage{url} 
\usepackage{multicol}
\usepackage{pslatex}

\begin{document} 

\title{Spatio-Temporal view on the Topological Functioning Model}

\author{Maria Spichkova} 

\institute{School of Computing Technologies, RMIT University, Melbourne, Australia\\
\email{maria.spichkova@rmit.edu.au}\\
%\url{http://www.springer.com/gp/computer-science/lncs}  
}
\maketitle

\abstract{ 
The spatial and temporal aspects of system properties are crucial for many types of systems. 
In this short paper, we present a TopFunST framework to analyse topological dependencies among features of the system, covering also spatial and temporal aspects. 
TopFunM is based on an extended version of the language of  Topological Functioning Models (TFM). 
The TFM language describes the topological relationships among functional features of the system, which allows not only to analyse system functionality, as well as feature interactions and functioning cycle structures, but spatial and temporal aspects haven't been covered yet. This paper presents a solution to this problem.  
}

% \begin{CCSXML}
% <ccs2012> 
% <concept>
% <concept_id>10010147.10010341.10010342</concept_id>
% <concept_desc>Computing methodologies~Model development and analysis</concept_desc>
% <concept_significance>300</concept_significance>
% </concept>
% <concept>
% <concept_id>10011007.10011006.10011039</concept_id>
% <concept_desc>Software and its engineering~Formal language definitions</concept_desc>
% <concept_significance>300</concept_significance>
% </concept> 
% </ccs2012>
% \end{CCSXML}

% \ccsdesc[500]{Computing methodologies~Model development and analysis}
% \ccsdesc[500]{Software and its engineering~Formal language definitions}

\newcommand{\epar}[1]{``#1''}
\newcommand{\focust}{\textsc{Focus}$^{ST}$}
\newcommand{\Focus}{\textsc{Focus}}
\newcommand{\te}[2]{\ensuremath{#1^{\left\lceil{#2}\right\rceil}}}
\newcommand{\Nat}{\ensuremath{\mathbb{N}}}
\newcommand{\tlangle}{\text{\textlbrackdbl}}
\newcommand{\trangle}{\text{\textrbrackdbl}}
\newcommand{\ntlangle}{{\normalfont\text{\textlbrackdbl}}}
\newcommand{\ntrangle}{{\normalfont\text{\textrbrackdbl}}}
\newcommand{\nif}{\textsf{if\ }}
\newcommand{\nfi}{\textsf{fi\ }}
\newcommand{\nelse}{\textsf{else\ }}
\newcommand{\nthen}{\textsf{then\ }}

% \onecolumn \maketitle \normalsize \setcounter{footnote}{0} \vfill

%====================================
\section{\uppercase{Introduction}}
\label{sec:introduction}
 
Model-driven software development (MDSE) provides many benefits for the software engineers: 
from the opportunities of early analysis of the core properties of the system to the application of the model-based testing approaches. 
However, success  and effectiveness of MDSE strongly depends on the quality and the appropriateness of the model.  
%An incomplete, inaccurate or inconsistent  
Generally,  a model is an abstraction of the system, which is created to focus the  system properties that have to be analysed.
at the corresponding development stage. Thus, the goal of the model is not to present the system on the corresponding level of abstraction, while hiding  the complexity of a system and its environment \cite{ahr2016enase}. However, if the critical information is missed in this abstraction, the model becomes useless or even dangerous: the engineers might expect system properties that do not reflect the actual system-under-development.  
On the other hand, if the model is not abstract enough, it might become unreadable.
The understandability and readability of the model is as crucial as accuracy and predictability.

In this short paper, we  analyse the topological dependencies among features of the system. 
For these purposes, we extended the language of  Topological Functioning Models (TFM) to spatio-temporal aspects. 
A TFM
describes the topological relationships among functional features of the system, which allows to analyse system functionality and functioning cycle structures, cf. \cite{osis2017topological}.
The input to TFM is an informal specification of the system (e.g., functional requirements specification). The model itself, in the terms of graph theory, is a directed graph, which nodes represent functional features of the system, while directed edges represent their topological relationships. % (more precisely, cause-and-effect relations between functional features). 
A number of approaches focused on the application of TFM to UML, e.g., \cite{donins2011towards,vslihte2010implementing}. 
However, all the above approaches did not cover any spatio-temporal aspects. Our work covers this limitation of TFM. 
 
%\textbf{Contributions:} 
We introduce a TopFunST framework that is based on an extended version of the the language of TFMs. 
The aim of TopFunST is to analyse topological dependencies among features of the system (which components becoming sp-objects), covering also spatial and temporal aspects.

% %=======================================
\section{\uppercase{Background}}
\label{sec:background}

The spatio-temporal aspects of the system properties play an important role. Thus,
the appropriate models of this kind of systems have to represent spatio-temporal aspects, allowing for both formal specification and verification. 
%In our earlier work we introduced 
\focust\ framework \cite{spichkova2014modeling,spichkova2016spatio} 
facilitates the development of comprehensible specifications, demonstrating its suitability for the application of the specification and proof methodology outlined in~\cite{spichkova,spichkova2013we}.
The \focust\ language was inspired by \Focus~\cite{focus},
a framework for formal specification and development of interactive systems. 
The specifications in both languages are based on the concept of \emph{streams}. 
The \focust\ specifications are a special form of timed automata that we refer to as \emph{Timed State Transition Diagrams} (TSTDs).

A TSTD can be described in both diagram and textual form. 
An input action for a TSTD can be specified as a set of current time intervals of the input streams of the system.
The output action can be specified as a set of corresponding time intervals of the output streams. 
In addition to its capability to represent timing properties within the language, the framework also enables the definition of special type of components specifying real objects  that are able to physically change their location in space., so-called 
\emph{sp-objects}, which provides an appropriate basis to create a topological functioning model covering spatio-temporal properties. 

While our current work is based on TSDT and TFM, there are also many other approaches aiming to specify spatio-temporal properties. For example, Pek at al.~\cite{pek2023spatial} proposed a spatio-temporal framework SpaTiaL that unifies the specification, monitoring, and planning of object-oriented robotic tasks in a robot-agnostic fashion.  
There were research studies on robot motion planning and spatio-temporal robotics planning conducted by Barbosa et al.~\cite{barbosa2021formal} and Santos et al.~ \cite{santos2017spatio}.
Yan et al.~\cite{yan2024spatio} proposed to conduct spatio-temporal analysis using the Signal Temporal Logic that focuses on the time series signal systems. 
Viseras et al.~\cite{viseras2020distributed} proposed an approach for multi-robot information gathering under spatio-temporal constraints.

%=======================================
\section{\uppercase{Formal Model}}
\label{sec:formal}

TopFunST extends the TFM language by spatio-temporal aspects as well as by the mappings to the approaches currently applied in industry.
The core artefacts in TopFunST to model the system behaviour are functional features.

We define a functional feature as a tuple $F$:
\[
F = \langle Id, A, R, Obj, Pre, Post, Prov, Exec, D, T, Loc\rangle
\]
where
\begin{itemize}
\item 
$Id$ is an action identifier/label;
\item 
$A$ is an action linked to an sp-object $Obj$ that performs this action,; %or is influenced by it (e.g., receives results of this action);
\item 
$R$ is a set of object influenced by the action;
\item 
$Pre$ is a set of preconditions  $\{pr_1,  \dots, pr_m\}$ of the action;
\item 
$Post$ is a set of postconditions  $\{ps_1, \dots, ps_m\}$ of the action;
\item
$Prov$  is a set of components that provide an
action with a set of certain objects;
\item
$Exec$ is a set of components that enact a concrete action.
\item
$D$ is a pair $(Dmin, Dmax)$ representing the minimal and the maximal duration of the action (similar to the idea of the Best and Worst Case Execution Times, used in the analysis of timing behaviour of real-time systems);
\item
$T$ represents timing constraints, e.g., how often the action should be repeated, in the case it should be performed in cycles;
\item
$Loc$ represents spatial aspects/constraints of the action.
\end{itemize}

In terms of requirements engineering, a functional feature can be seen as an event within a use case. 
A use case  can be defined as a tupel $UseCase$:
\[
UseCase  = \langle Act, UCPre, Events, Alt \rangle
\]
where
\begin{itemize}
\item 
$Act$ is the set of actors (external to the specified system);
\item 
$UCPre$ is the set of preconditions for $UseCase$;
\item
$Events$ is an ordered list presenting the flow of events $[E_1, \dots, E_n]$, 
where each event $E_i$ is specified by its number/id $i$ and the textual description of the event;
\item
$Alt$ a list presenting alternative flows, where the alternations refer to the corresponding event numbers/ids within the $Events$ list;
\end{itemize}
Thus, $Events$ and $Alt$ correspond to the subset of all possible paths within the directed graph, which specifies all CERs among features, and $UCPre$ specifies all the initially required preconditions of the corresponding actions.

In TFM, the Cause \& Effect Relations (CER) are represented as directed edges between nodes of a directed graph. %, which are oriented from a cause node to an effect node.
These relations  can also be specified in the form of an adjacency (incident) $n \times n$ matrix, 
where $n$ is the number of functional features represented as nodes in the topological space. 
If there is an edge from some node $x$ to some node $y$, 
then the matrix element $m_{x,y}$ is 1, otherwise, it is 0. 
This allows efficient analysis of the model to identify subgraphs, which is crucial for the analysis of the subsystems.
Moreover, this provides a solid basis for the analysis of the unwanted feature interactions, to exclude the case that  execution an action $a$ might lead to an unwanted execution of action $b$.

% %=======================================
% \section{\uppercase{Example: Intelligent Transportation Systems Scenario}}
% \label{sec:application}

\textbf{Example:}
Let us illustrate the core features of TopFunST using a  scenario with the autonomous robots working together to safely lift a carriage and deliver it to from location $L_1$ to location $L_2$.    
The set of sp-objects consists of three elements: two robots $R_1$ and $R_2$, as well as a carriage $C$: 
\[\{ R_1, R_2, C \}.\]
An (external) actor for this system of two robots will be a crane that places the carriage to $L_1$.

\begin{table*}[ht!]
\caption{System specification using TopFunST}
\begin{center}
{\scriptsize
\begin{tabular}{|l|l|l|l|l|l|l|l|l|l|l|}
\hline
\hline
Id & A & R & Obj & Pre & Post & Prov & Exec & D & T & Loc
\\
\hline
\hline
a & Moving & $R_i$, $[C]$  & $R_i$ & $MovementPossible$ & $LocationUpdated$ &  &  & $D_{Rmovement}$ & $TC$ & $L_1$
\\
\hline
b & Moving & $R_1, R_2, C$  & $C$ & $CarryingPossible$ & $LocationUpdated$ &  &  & $D_{Rmovement}$ & $TC$ & $L_1, L2, L3$
\\
\hline
c & Carrying & $R_1, R_2, C$  & $R_1, R_2$ & $CarryingPossible$ & $LocationUpdated$ &  &  & $D_{Rmovement}$ & $TC$ & $L_1, L2, L3$
\\
\hline
d & OpenGrippers & $R_i$ & $R_i$ & $GrippersClosed$ & $GrippersOpen$ &   &   & $D_{Gopening}$ & T & $L_i$
\\
\hline
e & CloseGrippers & $R_i$ & $R_i$ &  $GrippersOpen$ &  $GrippersClosed$   &   &   & $D_{Gopening}$ & T & $L_i$
\\
\hline
\hline
\end{tabular}
}
\end{center}
\label{tab:features}
\end{table*}% 

Table \ref{tab:features} presents the functional features of the system, where $i \in \{1,2\}$ and $[C]$ denotes an optional (constraint-dependent) case.
$TC$ present the timing constraint.
$D_{Rmovement}$ describes the minimal and maximal acceptable speed of the movement.

%=====================================
%\bibliographystyle{splncs04} 
%\bibliography{biblio}
% \bibliographystyle{apalike}
% {\small
% \bibliography{biblio}
% }

\end{document}